\documentclass{ws-procs961x669}  % to draw border line around text area
\newcommand{\sT}{{\scriptscriptstyle T}}
\renewcommand{\d}{\mathrm{d}}
\newcommand{\xB}{x_{\scriptscriptstyle B}}

\begin{document}
 
\title{Gluon TMDs and Opportunities at an EIC}

\author{Cristian Pisano}

\address{Dipartimento di Fisica, Universit\`a di Cagliari, and INFN, Sezione di Cagliari\\
 Cittadella Universitaria, I-09042 Monserrato (CA), Italy\\
$^*$E-mail: cristian.pisano@ca.infn.it}
%www.university\_name.edu}

\begin{abstract}
 
We show how transverse momentum dependent gluon distributions  could be probed at a future Electron-Ion Collider through the analysis of transverse momentum spectra and azimuthal asymmetries in the inclusive electroproduction of $J/\psi$ and $\Upsilon$ mesons. The maximum values of these asymmetries, obtained in a model-independent way by imposing the positivity bounds of the polarized gluon distributions, suggest the feasibility of the proposed measurements.

\end{abstract}

\keywords{Gluon distributions; quarkonium states; proton structure.}

\bodymatter

\section{Introduction}

Transverse momentum dependent distributions (TMDs) of unpolarized and polarized gluons represent a novel way of exploring the structure of the proton. They encode information on the transverse motion of partons and the correlation between spin and partonic transverse momenta~\cite{Mulders:2000sh}, providing a more complete description than the usual parton distribution functions, integrated over transverse momenta. Gluon TMDs are also of great interest because of their intrinsic process dependence, due to their gauge link structure: an unambiguous verification of this property will represent an important confirmation of our actual knowledge of the TMD formalism and nonperturbative QCD in general. 

At present, almost nothing is experimentally known about gluon TMDs. However, many proposals have been put forward to access them, mainly by looking at transverse momentum distributions and azimuthal asymmetries for bound or open heavy-quark pair production, both in lepton-proton and in proton-proton collisions. In particular,  the process $e\, p \to e^\prime \,Q \,\overline{Q}\, X$, with $Q$ being either a charm or a bottom quark, has been proposed as a tool to probe gluon TMDs at the future Electron-Ion Collider (EIC), which will be based in the US. In a series of papers~\cite{Boer:2010zf,Pisano:2013cya,Boer:2016fqd} it has been shown that five different gluon TMDs contribute to the unpolarized and transversely polarized cross sections, through specific azimuthal modulations. The measurements of properly defined azimuthal moments would allow to single out and  extract all the distributions.  Moreover, attention has been paid to the small-$x$ behavior of these observables and to their  process dependence properties, by relating them to analogous observables defined in proton-proton collisions.  

Along the same lines, in the following we describe a complementary analysis for the case in which two heavy quarks produced in a semi-inclusive deep inelastic scattering  process (SIDIS) form a bound state, specifically a $J/\psi$ or a $\Upsilon$ meson~\cite{Bacchetta:2018ivt,Mukherjee:2016qxa}.

\section{$J/\psi$ and $\Upsilon$ production in SIDIS}
\begin{figure}[t]
\begin{centering}
\includegraphics[trim={2cm 23cm 0cm 3cm},clip,scale=1]{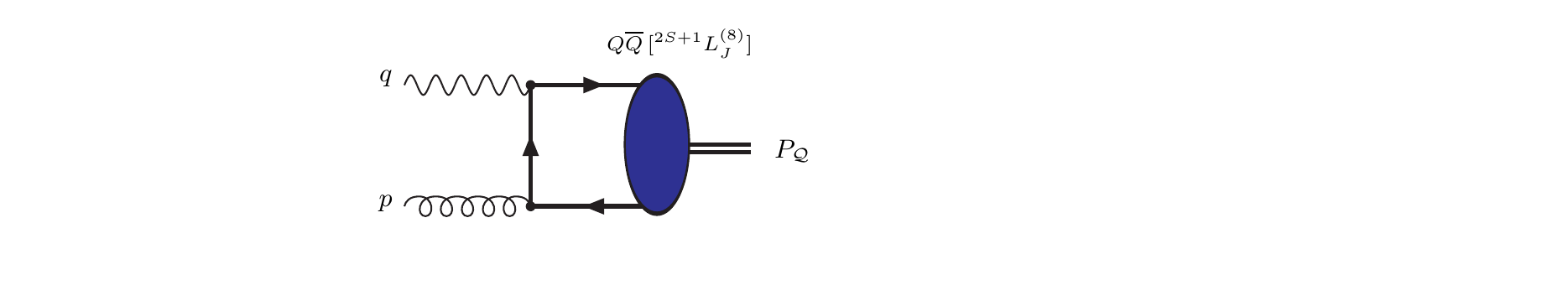}
\par\end{centering}
\caption{LO Feynman diagram for the partonic process $\gamma^* (q) \,+ \, g(p)\to {\cal Q} (P_{\cal Q})$, with ${\cal Q} = J/\psi$ or $\Upsilon$. The crossed diagram, in which the directions of the arrows are reversed, has to be considered as well. The $^1S_0^{(8)}$ and $^3P^{(8)}_{J}$ configurations, with $J=0,1,2$, are the only nonzero contributions. }
\label{fig:fd-lo}
\end{figure}

We consider the process $e\, p^{\uparrow} \to e \,J/\psi \,(\Upsilon)\, X$, where the initial proton is polarized with polarization vector $\bm S$ and $q_\sT \equiv \vert \bm q_\sT\vert$, the quarkonium momentum transverse w.r.t.\ the lepton plane, is small compared to its mass $M_{\cal Q}$ and to the virtuality  $Q$  of the photon exchanged in the reaction.
In a reference frame in which the proton and the photon move along the $\hat z$-axis and the azimuthal angles are measured w.r.t.\ to the lepton plane, the cross section has the following structure
\begin{equation}
\frac{\d\sigma}
{\\d y\,\d\xB\,\d^2\bm{q}_{\sT}} \equiv \d\sigma (\phi_S, \phi_\sT) =    \d\sigma^U(\phi_\sT)  +  \d\sigma^T (\phi_S, \phi_\sT)  \,,
\label{eq:cs}
\end{equation}
where $y$ and $\xB$ are the inelasticity and Bjorken variables, respectively, while $\phi_S$ and $\phi_\sT$ denote the azimuthal angles 
of the transverse vectors $\bm S_\sT$ and $\bm q_\sT$. The unpolarized cross section reads
\begin{align}
\d\sigma^U
  & =  {{\cal N}}\, \bigg [ A^U  f_1^g (x, \bm q_\sT^2 )+  \frac{\bm q_\sT^2}{M_p^2}\, B^U\, h_1^{\perp\, g} (x, \bm q_\sT^2 ) \cos 2 \phi_\sT  \bigg ] \,,
\label{eq:csU}
\end{align}
with ${\cal N} =   4 \pi^2 {{\alpha^2 \alpha_se_Q^2}}/[{ y\,  Q^2\, M_{\cal Q}(M_{\cal Q}^2+Q^2)}]$, $e_Q$ being the fractional electric charge of the quark $Q$, and $M_p$ the proton mass. Moreover,  $f_1^g $ is the unpolarized gluon TMD distribution and  $h_1^{\perp\,g} $ is the distribution of linearly polarized gluons inside an unpolarized proton. The amplitudes $A^U$ and $B^U$ can be calculated within the framework of nonrelativistic QCD (NRQCD). As depicted in Fig.~\ref{fig:fd-lo}, at leading order (LO) in the strong coupling constant $\alpha_s$, the partonic subprocess that contributes to $J/\psi$ production is $\gamma^*g \to Q \overline Q  [ ^{2S+1}L_J^{(8)} ] $. The spectroscopic notation indicates that the $Q \overline Q$ pair forms a bound state with  spin $S$, orbital angular momentum $L$, total angular momentum $J$ and color octet configuration (8). We find that $A^U$ and $B^U$ depend on the relevant long-distance matrix elements (LDMEs) $\langle 0 \vert {\cal O}_8^{J/\psi} (^1 S_0)\vert 0 \rangle$ and $\langle 0 \vert {\cal O}_8^{J/\psi} (^3 P_J)\vert 0 \rangle$, with $J=0,1,2$. Similarly, we obtain
%%%%%%%%%%%%%%%%%%%%%%%%%%%%%%%%%%%%%%%%%%%%%%%%%%%%%%%%%%%%%%%%%
%%%%%%%%%%%%%%%%%%%%%%%%%%%%%%%%%%%%%%%%%%%%%%%%%%%%%%%%%%%%%%%%%
\begin{align}
\d\sigma^T & =   {\cal N}\,\vert \bm S_\sT\vert \,\frac{\vert\bm q_\sT\vert }{M_p} \bigg \{A^T f_{1T}^{\perp\,g} (x, \bm q_\sT^2 ) \sin(\phi_S -\phi_\sT)  + B ^T \,\left [ h_{1}^{g} (x, \bm q_\sT^2 )\, \sin(\phi_S+\phi_\sT)  \right .
\nonumber \\ 
& \qquad \qquad \qquad \qquad -\left .  \,\frac{\bm q_\sT^2}{2 M_p^2}\, h_{1\sT}^{\perp\,g} (x, \bm q_\sT^2 )\, \sin (\phi_S - 3 \phi_\sT)  \right ] \bigg \} \, ,
\label{eq:csT}
\end{align}
where $f_{1T}^{\perp\,g}$ is the gluon Sivers function, while $h_{1}^{g}$ and $h_{1\sT}^{\perp\,g}$ are chiral-even distributions of linearly polarized gluons inside a transversely polarized proton. They are all $T$-odd, {\it i.e.}\ they would be zero in absence of initial or final state interactions. Therefore, they contribute to the cross section only because the underlying production mechanism is the color-octet one. 

\begin{figure}[t]
\begin{centering}
\hspace*{-0.5cm}
\includegraphics[clip,scale=.65]{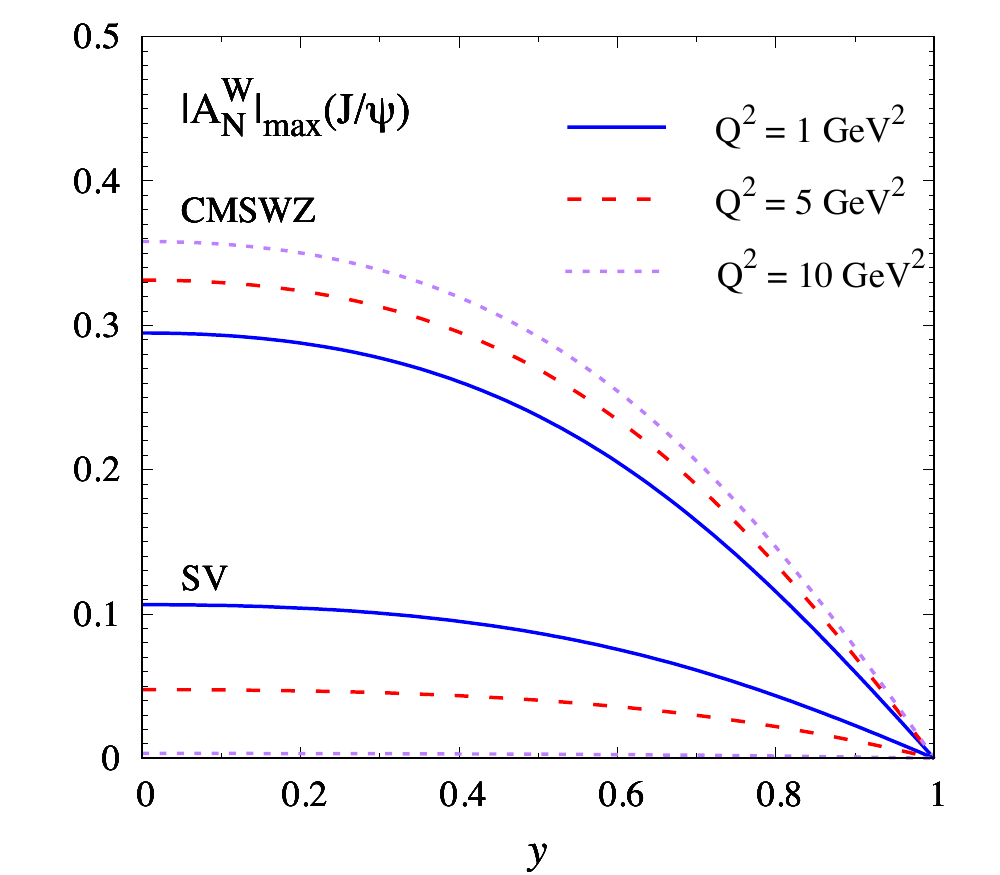}\includegraphics[clip,scale=.65]{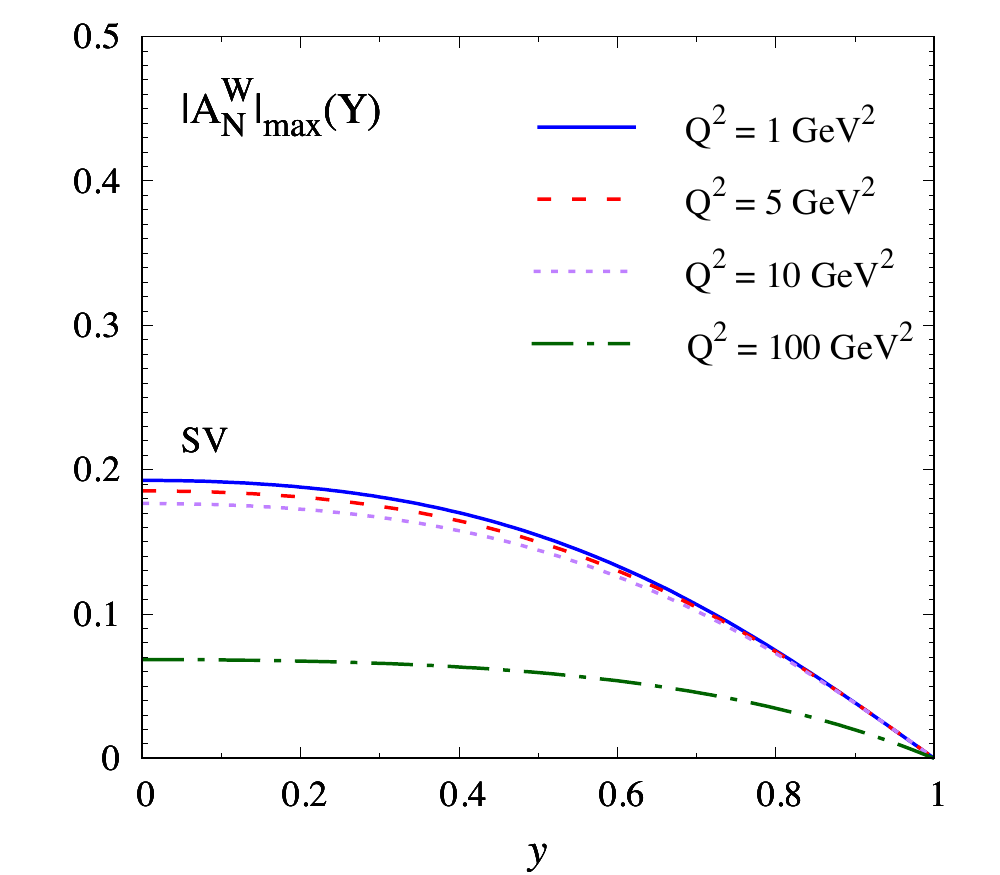}
\par\end{centering}
\caption{ Maximal $\langle \cos 2\phi_\sT\rangle$ and  $A_N^W$  asymmetries, with $W =  \sin (\phi_S +\phi_\sT),\, \sin (\phi_S - 3 \phi_\sT)$, for $J/\psi$ (left panel) and  $\Upsilon$ (right panel) production in SIDIS. The labels SV and CMSWZ refer to the adopted LDME sets~\cite{Chao:2012iv,Sharma:2012dy}.}
\label{fig:cos2phisat} 
\end{figure}

In order to single out the different azimuthal modulations, each one corresponding to a different gluon TMD, we introduce the following azimuthal moments
\begin{align}
A^{W(\phi_S,\phi_\sT)} & \equiv 2\,\frac {\int   \d \phi_S \, \d \phi_\sT \, W(\phi_S,\phi_\sT)\,\d\sigma (\phi_S,\,\phi_\sT)}{\int  \d \phi_S\,  \d \phi_\sT \,\d\sigma (\phi_S, \phi_\sT)}\,.
\label{eq:mom}
\end{align}
Furthermore, by taking $W= \cos2\phi_\sT$, we define $A^{\cos 2\phi_\sT} \equiv 2 \langle \cos 2\phi_\sT \rangle $.  The maximum values of such asymmetries, obtained from the positivity bounds of the TMDs, are presented in Fig.~\ref{fig:cos2phisat} in a kinematic region accessible at the EIC. They turn out to be sizable, but depend very strongly on the specific set of the adopted LDMEs. We point out that a measurement of the ratios
\begin{align}
\frac{A^{\cos 2\phi_\sT }}{A^{\sin(\phi_S+\phi_\sT)}}  \!= \! \frac{\bm q_\sT^2}{M_p^2}\, \frac{h_{1}^{\perp\,g}}{h_{1}^{g}}\,, \,
\frac{A^{\sin(\phi_S-3\phi_\sT)}}{A^{\cos 2\phi_\sT }} \!=\!- \frac{\vert \bm q_\sT\vert}{2 M_p}\, \frac{h_{1 \sT}^{\perp\,g}}{h_{1}^{\perp\,g}}\,, \,
\frac{A^{\sin(\phi_S-3\phi_\sT)}}{A^{\sin(\phi_S+\phi_\sT)}}\!  =\!- \frac{\bm q_\sT^2}{2 M_p^2}\, \frac{h_{1T}^{\perp\,g}}{h_{1}^{g}}
\label{eq:ratioA3}
\end{align}
would directly probe the relative magnitude of the various TMDs, without any dependence on the LDMEs.

%%%%%%%%%%%%%%%%%%%%%%%%%%%%%%%%%%%%%%%%%%%%%%%%%%%%%%%%%%%
%%%%%%%%%%%%%%%%%%%%%%%%%%%%%%%%%%%%%%%%%%%%%%%%%%%%%%%%%%%%
%
\section{Conclusions}

We have presented the LO expressions of the azimuthal asymmetries for $J/\psi$ and $\Upsilon$ production in SIDIS processes, obtained in the framework of NRQCD and assuming TMD factorization. Our results are valid when the transverse momentum of the quarkonium state is much smaller than its invariant mass, and can be used to gather information on gluon TMDs. To this aim, we have proposed the measurement of ratios of asymmetries in which the $\langle0\vert{\cal O}_{8}^{J/\psi}(^{1}S_{0})\vert0\rangle$ and $\langle0\vert{\cal O}_{8}^{J/\psi}(^{3}P_{0})\vert0\rangle$ long-distance matrix elements cancel out. Moreover, these asymmetries can shed light on the mechanism underlying quarkonium production in a totally novel way, for example if we compare them with the ones for $e\, p \to e^\prime \,Q \,\overline{Q}\, X$, at the same hard scale in order not to include TMD evolution~\cite{Bacchetta:2018ivt}. The method consists in the definition of other ratios in which the TMDs cancel out. Hence one would directly probe the two color-octet matrix elements,  which are still poorly known. 

 To conclude, we point out that the study of $J/\psi+$jet production at the EIC has been proposed very recently as an additional tool to access gluon TMDs~\cite{DAlesio:2019qpk}. In this case the soft scale is given by the total transverse momentum of the $J/\psi$+jet pair, required to be much smaller than its invariant mass. The advantage would be that, by varying the invariant mass of the pair, one could access a wide range of scales, having the opportunity to map out TMD evolution.

\bibliographystyle{ws-procs961x669}

\begin{thebibliography}{99}    
 %\cite{Mulders:2000sh}
\bibitem{Mulders:2000sh} 
  P.~J.~Mulders and J.~Rodrigues,
  Transverse momentum dependence in gluon distribution and fragmentation functions,
  {\it Phys. Rev. D} {\bf 63}, 094021 (2001)
%  doi:10.1103/PhysRevD.63.094021
   [hep-ph/0009343].
  %%CITATION = doi:10.1103/PhysRevD.63.094021;%%
 
%\cite{Boer:2010zf}
\bibitem{Boer:2010zf} 
  D.~Boer, S.~J.~Brodsky, P.~J.~Mulders and C.~Pisano,
 Direct probes of linearly polarized gluons inside unpolarized hadrons,
  {\it Phys.\ Rev.\ Lett.}\  {\bf 106}, 132001 (2011)
 % doi:10.1103/PhysRevLett.106.132001
  [arXiv:1011.4225 [hep-ph]].
  %%CITATION = doi:10.1103/PhysRevLett.106.132001;%% 
   
%\cite{Pisano:2013cya}
\bibitem{Pisano:2013cya} 
  C.~Pisano, D.~Boer, S.~J.~Brodsky, M.~G.~A.~Buffing and P.~J.~Mulders,
 Linear polarization of gluons and photons in unpolarized collider experiments,
  {\it JHEP} {\bf 1310}, 024 (2013)
%  doi:10.1007/JHEP10(2013)024
  [arXiv:1307.3417 [hep-ph]].
  %%CITATION = doi:10.1007/JHEP10(2013)024;%%

%\cite{Boer:2016fqd}
\bibitem{Boer:2016fqd} 
  D.~Boer, P.~J.~Mulders, C.~Pisano and J.~Zhou,
 Asymmetries in heavy quark pair and dijet Production at an EIC,
  {\it JHEP} {\bf 1608}, 001 (2016)
%  doi:10.1007/JHEP08(2016)001
  [arXiv:1605.07934 [hep-ph]].
  %%CITATION = doi:10.1007/JHEP08(2016)001;%%
  
  %\cite{Bacchetta:2018ivt}
\bibitem{Bacchetta:2018ivt} 
  A.~Bacchetta, D.~Boer, C.~Pisano and P.~Taels,
  Gluon TMDs and NRQCD matrix elements in $J/\psi$ production at an EIC,
  arXiv:1809.02056 [hep-ph].
  %%CITATION = ARXIV:1809.02056;%%

%\cite{Mukherjee:2016qxa}
\bibitem{Mukherjee:2016qxa} 
  A.~Mukherjee and S.~Rajesh,
  $J/\psi $ production in polarized and unpolarized $ep$ collision and Sivers and $\cos 2\phi $ asymmetries,  {\it Eur.\ Phys.\ J.\ C} {\bf 77}, 854 (2017)
%  doi:10.1140/epjc/s10052-017-5406-4
  [arXiv:1609.05596 [hep-ph]].

\bibitem{Chao:2012iv}
K-T.~Chao, Y-Q.~Ma, H-S.~Shao, K.~Wang and Y-J.~Zhang, $J/\psi$ polarization at hadron colliders in nonrelativistic QCD
{\it Phys.\ Rev.\ Lett.}\ {\bf108}, 242004 (2012) [arXiv:1201.2675 [hep-ph]].

\bibitem{Sharma:2012dy}
R.~Sharma and I.~Vitev, High transverse momentum quarkonium production and dissociation in heavy ion collisions,
{\it Phys.\ Rev.\ C} {\bf87},  044905  (2013) [arXiv:1203.0329 [hep-ph]]. 
 
 %\cite{DAlesio:2019qpk}
\bibitem{DAlesio:2019qpk} 
  U.~D'Alesio, F.~Murgia, C.~Pisano and P.~Taels,
  Azimuthal asymmetries in semi-inclusive $J/\psi\,+\,\mathrm{jet}$ production at an EIC, 
   {\it  Phys.\ Rev.\ D} {\bf 100}, 094016 (2019)
 [arXiv:1908.00446 [hep-ph]].
  %%CITATION = ARXIV:1908.00446;%%
    
\end{thebibliography}

\end{document}